\documentclass[twocolumn,aps,prl,superscriptaddress]{revtex4}
\usepackage{amssymb}
\usepackage{amsmath,bm}
\usepackage{graphicx}
\usepackage[normalem]{ulem}
\usepackage{multirow}
\usepackage[colorlinks,linkcolor=blue,urlcolor=blue,citecolor=blue]{hyperref}
\usepackage{lipsum}
\usepackage[usenames, dvipsnames]{xcolor}
\usepackage{tensor}
\usepackage{isotope}
\usepackage{amsmath}
\allowdisplaybreaks[4]
 
\renewcommand{\sout}{\bgroup \color{red} \ULdepth=-.5ex \ULset}
 
\newcommand{\pdt}{$tp/d^2~$} 
\begin{document}

\title{Spinodal  Enhancement of Light Nuclei Yield Ratio in Relativistic Heavy Ion Collisions}
\author{Kai-Jia Sun}
\email{kjsun@tamu.edu}
\affiliation{Cyclotron Institute and Department of Physics and Astronomy, Texas A\&M University, College Station, Texas 77843, USA}

\author{Wen-Hao Zhou} 
\affiliation{Faculty of Science, Xi’an Aeronautical Institute, Xi’an 710077, China}

\author{Lie-Wen Chen}
\email{lwchen@sjtu.edu.cn}
\affiliation{School of Physics and Astronomy, Shanghai Key Laboratory for Particle Physics and Cosmology, and Key Laboratory for Particle Astrophysics and Cosmology (MOE), Shanghai Jiao Tong University, Shanghai 200240, China}
   
\author{Che Ming Ko}
\email{ko@comp.tamu.edu}
\affiliation{Cyclotron Institute and Department of Physics and Astronomy, Texas A\&M University, College Station, Texas 77843, USA}

\author{Feng Li}
\email{fengli@lzu.edu.cn}
\affiliation{School of Physical Science and Technology, Lanzhou University, Lanzhou, Gansu, 073000, China \\
Lanzhou Center for Theoretical Physics, Key Laboratory of Theoretical Physics of Gansu Province, and Frontiers Science Center for Rare Isotopes, Lanzhou University, Lanzhou 730000, China} 

\author{Rui Wang}  
\affiliation{Key Laboratory of Nuclear Physics and Ion-beam Application~(MOE), Institute of Modern Physics, Fudan University, Shanghai $200433$, China}

\author{Jun Xu}
\email{xujun@zjlab.org.cn}
\thanks{\\Kai-Jia Sun and Wen-Hao Zhou contributed equally to this work.}
\affiliation{Shanghai Advanced Research Institute, Chinese Academy of Sciences, Shanghai 201210, China }

\date{\today}

\begin{abstract}
Using a relativistic transport model  to describe the evolution of the quantum chromodynamic matter produced in Au+Au collisions at $\sqrt{s_{NN}}=3-200$ GeV, we study the effect of a first-order phase transition in the equation of state of this matter on the yield ratio $N_tN_p/ N_d^2$ ($tp/d^2$) of produced proton ($p$), deuteron ($d$), and triton ($t$). We find that the large density inhomogeneities generated by the spinodal instability during the first-order phase transition can survive the fast expansion of the subsequent hadronic matter and lead to an enhanced $tp/d^2$ in central collisions at $\sqrt{s_{NN}}=3-5$~GeV as seen in the experiments by the STAR Collaboration and the E864 Collaboration.  However, this enhancement subsides  with increasing collision centrality, and the resulting almost flat centrality dependence of $tp/d^2$ at $\sqrt{s_{NN}}=3$ GeV can also be used as a signal for the first-order phase transition. 
\end{abstract}
 
\maketitle

\emph{Introduction.}{\bf ---}
The search for a first-order phase transition and the critical point (CP) in the phase diagram of Quantum Chromodynamics (QCD) is  the central goal of the beam energy scan (BES) program in relativistic heavy ion collisions~\cite{Bzdak:2019pkr,Luo:2017faz}. The  transition from the quark-gluon plasma (QGP) to the hadronic matter  at vanishing baryon density $\rho_B$ or baryon chemical potential $\mu_B$ is known to be a smooth crossover~\cite{Aok06,Baz12}.  This phase transition is, however, expected to  change to a first-order one at finite $\mu_B$~\cite{Asakawa:1989bq,Fukushima:2010bq,Fu:2019hdw},  resulting in a critical (end)point on the first-order phase transition line in the $\mu_B$-$T$ plane of the QCD phase diagram.

In contrast to the critical point, which is characterized by a divergent correlation length~\cite{Stephanov:1999zu,Stephanov:2008qz}, a first-order phase transition is accompanied by  a negative compressibility, giving rise to the so-called spinodal instabilitiy~\cite{Chomaz:2003dz,Randrup:2003mu}. The associated non-equilibrium dynamics can lead to inhomogeneities in the  density distribution of produced baryons~\cite{Sasaki:2007db,Steinheimer:2012gc}.
Results from studies using models based on the hydrodynamic approach~\cite{Steinheimer:2012gc,Steinheimer:2013gla,Herold:2013bi,Herold:2013qda} and the transport  approach~\cite{Li:2016uvu} have shown that the small initial density inhomogeneities in heavy ion collisions are amplified by the spinodal instability when the evolution trajectory of the produced matter traverses the spinodal region of the QCD phase diagram.  These enhanced density fluctuations can leave imprints on observables like the event-by-event fluctuations of conserved charges~\cite{Baym:1999up,Steinheimer:2019iso}, the direct flow of net baryons~\cite{Stoecker:2004qu,STAR:2014clz}, and the yields of dileptons~\cite{Kajantie:1986dh,Li:2016uvu,Rapp:2014hha,Seck:2020qbx},  as well as those of light nuclei~\cite{Steinheimer:2013gla,Sun:2017xrx,Sun:2018jhg,Shuryak:2018lgd,Shuryak:2019ikv,Sun:2020zxy}. 
 
\begin{figure*}[t]
  \centering
  \includegraphics[width=0.9\textwidth]{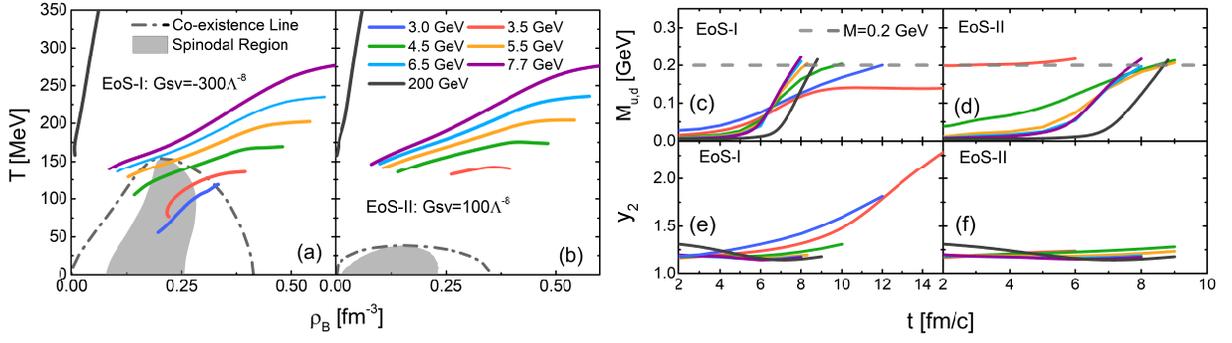}
  \caption{Left window: Evolution trajectories of the quark matter produced in central Au+Au collisions at $\sqrt{s_{NN}}=3-200$ GeV (solid lines) for EoS-I with $T_c=154$ MeV and EoS-II with $T_c=39$ MeV. The shaded regions denote the spinodal region and the dashed-dotted lines denote the co-existence lines. Right window: Time evolution of the in-medium quark mass $M_{u,d}$ (upper panels) and the scaled density moment $y_2$ (lower panels).}
  \label{pic:gsv}
\end{figure*}

For light nuclei production, studies based on the coalescence model have shown that the yield ratio $N_tN_p/ N_d^2$ ($tp/d^2$) would be enhanced if large nucleon density fluctuations are present at the late stage of hadronic matter expansion~\cite{Sun:2017xrx,Sun:2018jhg}.  Experimentally, the preliminary data from heavy ion collisions at both RHIC BES energies~\cite{Zhang:2020ewj} and SPS energies~\cite{Anticic:2016ckv} have shown possible enhancements and non-monotonic structures in the collision energy dependence of the ratio $tp/d^2$, which can not be described by  calculations based on transport or hybrid hydrodynamic  models without including the effects of spinodal instability and/or long-range correlations. Although it was shown in Ref.~\cite{Sun:2020pjz} that including the spinodal effect in a transport model with a first-order phase transition would lead to an enhancement of  $tp/d^2$, this study was based on a schematic fireball initial condition for heavy ion collisions at only one collision energy. A   systematic  study with realistic initial conditions and for various collision energies  is needed to extract from the experimental data the equation of state of produced QCD matter.

In the present study, we investigate the effect of spinodal instability in relativistic heavy ion collisions by developing a novel transport model based on an extended Nambu-Jona-Lasinio (NJL) model~\cite{Nam611,Nam612,Buballa:2003qv} that includes an eight-quark scalar-vector interaction~\cite{Sun:2020bbn} for the partonic matter. This extended NJL model~\cite{Sun:2020bbn} provides a flexible equation of state with  the critical temperature  $T_c$ ranging from around 40 MeV to 150 MeV, depending on the strength of the scalar-vector interaction,  which makes it possible to study realistically the effect of spinodal instability in Au+Au collisions at $\sqrt{s_{NN}}=3-200$ GeV. With  this new approach, we find a clear enhancement of \pdt due to the spinodal instability in heavy ion collisions at $\sqrt{s_{NN}}\approx 3-5$~GeV for an equation of state (EoS) with a critical temperature $T_c\gtrsim 80$ MeV.  Also, we find that such an enhancement leads to a flat centrality dependence of $tp/d^2$ at $\sqrt{s_{NN}}=3$ GeV, which can be  used as a signal  for the first-order phase transition.

\emph{Methods.}{\bf ---}
%2.1 framework
For the initial parton distribution in our transport model study, we take it from a multiphase transport (AMPT) model~\cite{Lin05}.  The time evolution of these partons in the phase space  is then described by their motions in the mean-field potentials and mutual scatterings.   After converting these quarks to hadrons via the spatial coalescence model in  AMPT, which retains the density fluctuations from the partonic matter, the subsequent hadronic rescatterings and decays are further modelled by a relativistic transport (ART) model~\cite{Li:1995pra} in AMPT.  At the kinetic freeze out when particles cease to scatter, the nucleon coalescence model is employed to evaluate the yields of light nuclei.

%2.2 partonic interaction
The Lagrangian density of the extended three-flavor NJL model with an eight-quark scalar-vector interactions is given by~\cite{Sun:2020bbn}
\begin{eqnarray}\label{eq:njlgsv}
 \mathcal{L}&=&\bar{\psi}(i\gamma^\mu \partial_\mu - \hat{m})\psi + G_{S}\sum_{a=0}^8 [(\bar{\psi}\lambda^a\psi)^2 + (\bar{\psi}i \gamma_5 \lambda^a \psi)^2] \notag \\
 &-&K\{[\text{det}[\bar{\psi}(1+\gamma_5)\psi] + \text{det}[\bar{\psi}(1-\gamma_5)\psi]\}\notag \\
 &+&G_{SV}\left\{\sum_{a=1}^3[(\bar{\psi}\lambda^a\psi)^2 + (\bar{\psi}i \gamma_5 \lambda^a \psi)^2]\right\}\notag \\
 &&\times\left\{\sum_{a=1}^3[(\bar{\psi}\gamma^\mu\lambda^a\psi)^2 + (\bar{\psi} \gamma_5 \gamma^\mu\lambda^a \psi)^2]\right\},
\end{eqnarray}
where  $\psi = (u,d,s)^T$ represents the 3-flavor quark fields, $\lambda ^a $(a=1,...,8) is the Gell-Mann matrices, and $\lambda^0 =\sqrt{2/3}I$, and $\hat{m}=\text{diag}(m_u,m_d,m_s)$ is the current quark mass matrix. In the above,  $G_{S}$ is the scalar coupling constant, $K$ is the coupling constant  for the Kobayashi-Maskawa-t'Hooft (KMT) interaction~\cite{tHooft:1976snw} that breaks the $U(1)_A$ symmetry with `det' denoting the determinant in flavor space~\cite{Buballa:2003qv}, and $G_{SV}$ is the coupling constant for the scalar-vector ($SV$) interaction~\cite{Sun:2020bbn}.  In the mean-field approximation, the $u$ quark mass is given by the gap equation $M_{u}= m_u - 4G_S\phi_u + 2K\phi_d\phi_s-2G_{SV}(\rho_u+\rho_d)^2(\phi_u+\phi_d)$ with $\phi_{i=u,d,s}$ and $\rho_i$ being the quark condensate and net-quark density, respectively. Similar expressions can be obtained for $d$ and $s$ quark masses.  In the vacuum  or in the chirally restored phase, where $\rho_i=0$ or $\phi_i=0$, the scalar-vector interaction does not affect the quark masses.   Because of the small current quark masses, the chiral symmetry is only approximately restored at high densities or temperatures. Since the location of the critical point in the QCD phase diagram can be easily changed by varying the value of $G_{SV}$, as shown in Ref.~\cite{Sun:2020bbn}, without affecting the properties of QCD vacuum, the present model is convenient and suitable for investigating the effects of spinodal instability in heavy ion collisions. Due to the non-renomalizability of the NJL model, a momentum cutoff $\Lambda$ is needed in evaluating the scalar and vector mean fields. For the parameters in  this extended NJL model, we use the values $m_u=m_d=5.5$~MeV, $m_s=140.7$~MeV, $G_S\Lambda^2=1.835$, $K\Lambda^5=12.36$, and  $\Lambda=602.3$~MeV~\cite{Lut92,Buballa:2003qv}. The partonic transport equation derived from the interaction Lagrangian in Eq.~(\ref{eq:njlgsv}) is solved by the test particle method~\cite{Sun:2020pjz,Wong:1982zzb}, with the details given in the supplemental material. 

\emph{Density fluctuations from the spinodal instability.}{\bf ---}
%3.1 phase trajectory and density fluctuation
The above transport model is used here to study central Au+Au collisions at $\sqrt{s_{NN}}=3-200$ GeV by firstly considering two equations of state with critical temperatures of $T_c=154$ MeV (EoS-I) and $T_c=39$ MeV (EoS-II), from setting  $G_{SV}=-300\Lambda^{-8}$ and $G_{SV}=100\Lambda^{-8}$, respectively. In panels (a) and (b) of Fig.~\ref{pic:gsv}, the dash-dotted lines depict the co-existence line of the low-density and high-density phases of equal pressure ($P$), $T$, and $\mu_B$.   In the shaded region, the quark matter has a negative compressibility  $\kappa_T^{-1}=\rho_B(\partial P/\partial \rho_B)_T<0$ and is unstable against phase decompositions~\cite{Chomaz:2003dz}.  Each point on the evolution trajectories (solid lines) in the $\rho_B-T$ plane of the phase diagram is obtained from the average temperature and baryon density of the matter in the central  volume of $4\times4\times4$~fm$^3$ weighted by the parton number in each cell of $1\times1\times1$~fm$^3$. For the temperature in each cell, it is determined by assuming that the energy density and the baryon number density in the rest frame of the cell are the same as those of a thermally equilibrated quark matter described by the present extended NJL model.  As shown in panel (a) of Fig.~\ref{pic:gsv}, for EoS-I the phase trajectories  traverse the spinodal region for collisions at $\sqrt{s_{NN}}<6.5$ GeV. In contrast, for EoS-II the phase trajectories  traverse only the crossover region at all collision energies, as shown in panel (b) of Fig.~\ref{pic:gsv}. 

As the temperature and density of the quark matter decrease during its expansion, the quark masses gradually increase from their current masses in the chirally restored  phase to larger masses in the chirally broken phase, as shown in Fig. 1 (c) and (d) for the time evolution of the  $u$ and $d$ quark masss $M_{u,d}$.  The only exception to this behavior of the quark masses is for EoS-II at $\sqrt{s_{NN}}<3.5$ GeV where the partonic stage is essentially absent because the chiral symmetry is  strongly broken throughout the evolution of the quark matter. Due to the lack of deconfinement transition in the NJL model, we assume in our study that the deconfinement transition coincides with the chiral phase transition as indicated in the lattice QCD calculations for QGP at zero baryon chemical potential~\cite{Karsch:2001cy}.  We thus convert quarks to hadrons once the chiral symmetry  becomes strongly broken when the average quark mass exceeds 200 MeV, which is about 55\% of its constituent mass in vacuum and corresponds to an energy density around 0.26 GeV/fm$^3$~\cite{Shen:2020jwv}.  With the hadronization carried out via the quark coalescence in the AMPT model, the spinodal effects are then retained during the  phase transition and the local baryon density fluctuation is also preserved. We further assume that the lifetime for the quark-hadron mixed-phase does not exceed $t_h=15$ fm/$c$, i.e., all partons are converted to hadrons before or at 15 fm/$c$. Prolonging the maximum value of $t_h$ only makes the spinodal effects more pronounced.   

To quantify the density fluctuations in the produced matter from heavy ion collisions, we use the second-order scaled density moment $y_2$~\cite{Steinheimer:2012gc}, defined as $y_2 =\overline{\rho^2}/\overline{ \rho}^2=[\int \text{d}{\bf x}\rho({\bf x})][\int \text{d}{\bf x}\rho^3(\bf{x})]/[\int \text{d}{\bf x}\rho^2({\bf x})]^{2}$, where $\rho({\bf x})$ denotes the matter density distribution in space. Figure 1 (e) and (f) clearly show that the $y_2$ can reach a value significantly greater than unity, which is expected from a uniform quark matter distribution, if the evolution trajectory of the produced matter enters the spinodal region of the phase diagram. After converting these quarks to hadrons via the spatial coalescence model in AMPT, the density fluctuations are smoothly converted from the partonic phase to the hadronic phase.  Due to the much larger expansion rate of the subsequent hadronic matter evolution than the hadron rescattering rate needed to maintain local equilibrium, the large density fluctuation is seen to survive  at the kinetic freeze-out~\cite{Sun:2020pjz}. 
   
\emph{Spinodal instability induced enhancement of $tp/d^2$.}{\bf ---}
Although the spatial density fluctuation/correlation in the matter produced from heavy ion collisions cannot be directly measured in experiments, it can affect the production of light (anti-)nuclei~\cite{Sun:2017xrx,Sun:2018jhg}. As shown in Refs.~\cite{Sun:2020zxy,Sun:2020pjz} based on the nucleon coalescence model, the yield ratio $tp/d^2$ is approximately given by $y_2/(2\sqrt{3})$ if the density-density correlation length is zero, which then leads to $tp/d^2\approx 0.29$ in the absence of density fluctuations when $y_2=1$. In the nucleon coalescence model for light nuclei production, which has been successfully used in high-energy nuclear collisions~\cite{Scheibl:1998tk,Sun:2018mqq,Bellini:2018epz,Zhao:2018lyf,Zhao:2021dka}, the formation probabilities of deuteron and triton from a proton-neutron pair and a proton-neutron-neutron triplet, respectively, in the produced hadronic matter at the kinetic freezeout are given by their Wigner functions. The latter are usually taken to be Gaussian, i.e., $W_\text{d}=8g_d\exp\left({-\frac{x^2}{\sigma_\text{d}^2}}-\sigma_\text{d}^2 p^2\right)$ and $W_\text{t}=8^2g_t\exp\left({-\frac{x^2}{\sigma_\text{t}^2}-\frac{\lambda^2}{\sigma_\text{t}^2}}-\sigma_\text{t}^2 p^2-\sigma_\text{t}^2 p_\lambda^2\right)$ with $g_d=3/4$ ($g_t=1/4$) being the statistical factor for spin 1/2 proton and neutron to from a spin 1 deuteron (spin 1/2 triton). The size parameters in the Wigner functions are $\sigma_\text{d}  = \sqrt{4/3}~r_{\text d}\approx 2.26$ fm and $\sigma_\text{t}=r_\text{t}=1.59$ fm, respectively, to reproduce the root-mean-squared matter radii of deuteron and triton~\cite{Sun:2020pjz,Ropke:2008qk}. The relative coordinates and momenta in $W_d$ and $W_t$ are defined as ${\bf x}=({\bf x}_1-{\bf x}_2)/\sqrt{2}$, ${\bf p}=({\bf p}_1-{\bf p}_2)/\sqrt{2}$, ${\boldsymbol\lambda}=({\bf x}_1+{\bf x}_2-2{\bf x}_3)/\sqrt{6}$, and $\quad{\bf p_\lambda}= ({\bf p}_1+{\bf p}_2-2{\bf p}_3)/\sqrt{6}$.

\begin{figure}[!t]
  \centering
  \includegraphics[width=0.416\textwidth]{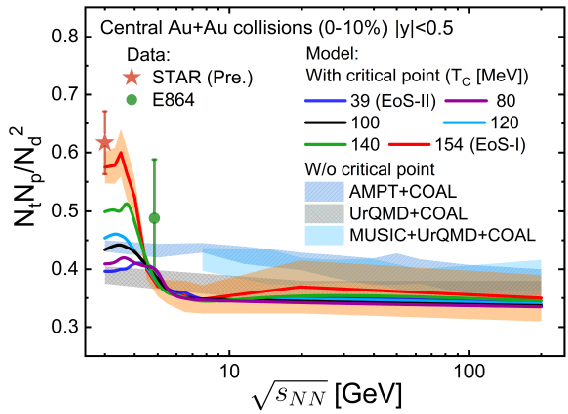}
  \caption{Collision energy dependence of the yield ratio \pdt from AMPT+COAL~\cite{Sun:2020uoj}, UrQMD+COAL~\cite{Bleicher:1999xi}, and MUSIC+UrQMD+COAL~\cite{Zhao:2021dka} as well as the present transport model study using various values for the critical temperature in the QCD phase diagram.  Experimental data are from the STAR Collaboration~\cite{Liu:2022} and the E864 Collaboration~\cite{E864:2000auv}.}
  \label{pic:pdt}
\end{figure}

Based on the nucleon phase-space distribution functions  at the kinetic freeze out from our transport model study of central Au+Au collisions, we have used the above coalescence model to study the collision energy dependence of the yield ratio $tp/d^2$.  Results obtained from using six different EoSs with critical temperatures of 39, 80, 100, 120, 140, 154 MeV by setting $G_{SV}=~$100, -27, -67, -119, -195, -300 $\Lambda^{-8}$, respectively, are shown in Fig.~\ref{pic:pdt} by solid lines.  For the case of EoS-II with $T_c=39$ MeV, the evolution  trajectory of the produced matter in the QCD phase diagram only traverses the crossover region.  In this case, the yield ratio  $tp/d^2$, shown by the blue line,  only shows a small increase at low collision energies, which might be due to the change in the volume of produced matter as the collision energy decreases. As $T_c$ increases, the evolution trajectory starts to enter the spinodal unstable region and the enhancement of \pdt at low energies becomes  increasingly prominent when $T_c>80$ MeV.  A maximum enhancement of \pdt of more than 50\% is achieved by using EoS-I with $T_c=154$ MeV. The slight decrease of \pdt at $\sqrt{s_{NN}}<3.5$ GeV is due to the shortening of the mixed-phase lifetime as seen in Fig.~\ref{pic:gsv}.  Also shown in Fig.~\ref{pic:pdt} by solid star and solid circle are the experimental data  from the STAR Collaboration~\cite{Liu:2022} and the E864 Collaboration~\cite{E864:2000auv}, respectively, and they are consistent with our theoretical results that include a first-order phase transition in an EoS with $T_c\approx$ 150 MeV.  In determining the experimental value of \pdt at $\sqrt{s_{NN}}=4.84~$GeV~\cite{E864:2000auv}, the yields of light nuclei of atomic number $A$  are obtained by fitting their transverse momentum ($p_T$) spectra at $p_T/A<0.3$ GeV with a blast-wave parameterization~\cite{Retiere:2003kf}.

Further shown in Fig.~\ref{pic:pdt} are the results from calculations using transport/hydrodynamics  approaches without a first-order phase transition in the equation of state together with the coalescence  model (COAL) for light nuclei production,  i.e., AMPT+COAL~\cite{Sun:2020uoj}, UrQMD+COAL~\cite{Bleicher:1999xi}, and MUSIC+UrQMD+COAL~\cite{Zhao:2021dka}.  These studies  all give an almost collision energy independent $tp/d^2$ and fail to explain the enhanced  \pdt at $\sqrt{s_{NN}}=3$ GeV.  Similar results have also been obtained in calculations based on more schematic coalescence models~\cite{Liu:2019nii,Hillmann:2021zgj}. Although allowing in the coalescence model a sequential production of light nuclei according to their decreasing atomic masses can lead to a strong enhancement of \pdt at $\sqrt{s_{NN}}\le$2.5 GeV, it has  negligible effects at $\sqrt{s_{NN}}\ge$3 GeV~\cite{Hillmann:2021zgj}. As to the (preliminary) data of \pdt in heavy ion collisions at higher collision energies of $\sqrt{s_{NN}}>6$ GeV, such as those at RHIC BES energies~\cite{Zhang:2020ewj} and SPS energies~\cite{Anticic:2016ckv}, the values of \pdt are affected by the weak decay corrections~\cite{Oliinychenko:2020znl,Zhao:2022xkz} to the proton yield,  which needs to be carefully calibrated before comparisons with model predictions are made, and they are therefore not shown in Fig.~\ref{pic:pdt}.

\begin{figure}[!t]
  \centering
  \includegraphics[width=0.42\textwidth]{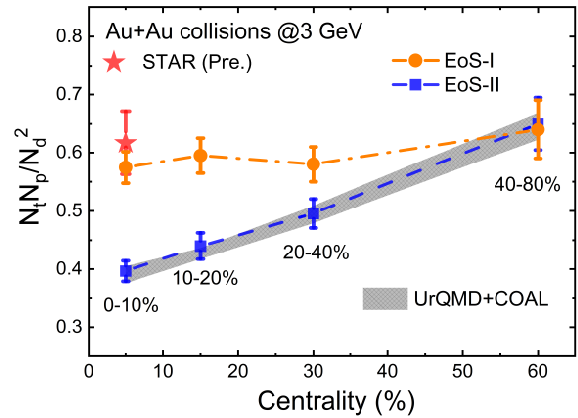} 
  \caption{Centrality dependence of the yield ratio \pdt in Au+Au collisions at $\sqrt{s_{NN}}=3$ GeV  for two different equations of state, with the experimental data taken from the STAR Collaboration~\cite{Liu:2022}. Also shown is the result from  the UrQMD+COAL model. }
  \label{pic:cen}
\end{figure}

Besides the collision energy dependence, the centrality dependence of the yield ratio $tp/d^2$  in heavy ion collisions can also reveal the occurrence of a first-order phase transition.  Because of increasingly shorter lifetime of the produced matter as the collision centrality increases, which limits the development of the spinodal instability during the phase transition, its induced enhancement of \pdt  is expected to subside in more peripheral collisions. This effect is clearly seen in Fig.~\ref{pic:cen}, which  depicts the centrality dependence of \pdt in Au+Au collisions at $\sqrt{s_{NN}}=3$ GeV for the two equations of state of EoS-I (solid circles) and EoS-II (solid squares). In the absence of  a first-order phase transition, i.e.,  EoS-II, the yield ratio \pdt is seen to  depend linearly on the collision centrality with a slope of 0.44$\pm$0.03, which agrees well with the results from UrQMD+COAL given by the gray band. A similar linear dependence but with a larger slope is obtained for AMPT+COAL (not shown). As shown in Ref.~\cite{Zhao:2021dka} from a study based on the MUSIC+UrQMD+COAL model and also the AMPT+COAL model, such a collision centrality dependence of \pdt is due to the decreasing size of the produced matter with increasing collision centrality.   On the other hand, using EoS-I with a first-order phase transition gives  an almost flat centrality dependence of $tp/d^2$ with a slope of $0.09\pm 0.05$ that is 5 times smaller, and this is because of the stronger enhancement of \pdt induced by the spinodal instability in more central collisions.  A flat centrality dependence of $tp/d^2$ can thus be used as a distinct signal for a first-order phase transition in relativistic heavy ion collisions and can be tested in RHIC FXT experiments.  

We note that the extracted value of $T_c\approx 150$~MeV  from our study  is larger than those predicted from theoretical approaches based on the functional renormalization method~\cite{Fu:2019hdw} and the Dyson-Schwinger equation~\cite{Gao:2015kea}. Since the inclusion of hadronic  mean-field potential, which is attractive for nucleons, in the hadronic transport is expected to help the survival of density fluctuations during the  hadronic matter expansion, it will further enhance the yield ratio $tp/d^2$, resulting in the need of an EoS with a smaller $T_c$  to reproduce the experimental data.  This effect requires a detailed study in the future. 
 
\emph{Conclusion.}{\bf ---} 
Motivated by the fact that finding the signals of a first-order phase transition in relativistic heavy ion collisions would indicate the existence of a critical point and also help determine its location in the QCD phase diagram, we have studied the effect of its associated spinodal instability. Using a novel transport model with a flexible equation of state of variable critical temperatures for relativistic heavy ion collisions, we have demonstrated that the large density inhomogeneities generated from the spinodal instability during the first-order QCD phase transition can  significantly enhance the yield ratio $tp/d^2$ at the kinetic freezeout. We have also found that the spinodal enhancement of \pdt is more prominent for EoS with higher critical temperatures and also in more central collisions.  In particular, we have found that the  spinodal enhancement of \pdt in central Au+Au collisions at $\sqrt{s_{NN}}\approx3-5$ GeV ($E_\text{lab}=3.85-12.4~$A~GeV) is in accordance with  the (preliminary) data from the STAR Collaboration and the E864 Collaboration.  Moreover, the almost flat centrality dependence of \pdt at $\sqrt{s_{NN}}=3$ GeV due to the spinodal instability serves as a cross-check for the occurrence of a first-order phase transition, and this can be tested by precise measurements in RHIC FXT experiments at $\sqrt{s_{NN}}=3-7.7$ GeV and in other experiments at CBM/FAIR, HADES/GSI, CEE/HIAF, NICA/MPD, NA61/SHINE, and J-PARC-HI.

\begin{acknowledgments}
K. J. Sun thanks Xiaofeng Luo, Dingwei Zhang, and Hui Liu for helpful discussions.
This work was supported in part by the US Department of Energy under Contract No.DE-SC0015266, the National Natural Science Foundation of China under Grant No. 11922514 and  No. 11625521, and National SKA Program of China No. 2020SKA0120300.
\end{acknowledgments}

%\bibliography{myref} 

\end{document}